\documentclass[a4paper]{jpconf}
\usepackage{graphicx}
\usepackage{epstopdf}
\begin{document}
\title{Dipole leakage and low CMB multipoles}

\author{Santanu Das and Tarun Souradeep}

\address{IUCAA, Post Bag 4, Ganeshkhind, Pune,411 007, India}

\ead{santanud@iucaa.ernet.in, tarun@iucaa.ernet.in}

\begin{abstract}
A number of studies of WMAP-7 have highlighted that the power at the low multipoles in CMB
power spectrum are lower than their theoretically predicted values. Angular 
correlation between the orientations of these low multipoles
have also been discovered. While these observations may have 
cosmological ramification, it is important to investigate possible observational
 artifacts that can mimic them. The CMB dipole which
is almost 550 times higher than the quadrupole can get leaked to the
higher multipoles due to the non-circular beam of the CMB experiment. In this paper
 an analytical method has been developed and simulations
are carried out to study the effect of the non-circular beam on power
leakage from the dipole. It has been shown that the small, but non-negligible power from the dipole can get
transferred to the quadrupole and the higher multipoles due to the
non-circular beam. Simulations have also been carried out for Planck
scan strategy and comparative results between WMAP and Planck have
 been presented in the paper.
\end{abstract}

\section{Introduction}
The standard model of cosmology emerging from recent observations is a remarkable success of theoretical
physics. It can explain the cosmological observations up
to an extremely high precision using a handful set of parameters.
However, there are some effects that seem to be anomalous in the standard cosmological
model. One of such observational fact is the power anomaly at the
low multipoles of the CMB power spectrum. It has been seen that
the power at the low multipoles are lower then their theoretical predictions
and there are possibly correlation between the orientations of the
low multipoles. Many researchers have tried to explain the phenomenon
 such as in \cite{Moss-2010,Feng2003}, but the phenomenon is not satisfactorily explained. In this paper we have analyzed the effect
of the non-circular beam as origin of CMB anomalies at low multipoles.

In most of the CMB experiments such as WMAP,
the beam shape of the detectors are non-circular about the pointing direction. However in the data
analysis techniques the beam is assumed to be circularly symmetric. The CMB dipole is almost $550$
times stronger than the quadrupole. Therefore due to the non-circularity
of the beam some power from dipole may leak to the quadrupole
and immediate higher multipoles. However, assuming a circular beam in the
data analysis technique, this leakage of the power from dipole will not be accounted for. 
Therefore, the contribution of this effect will contaminate the resultant map, generated by this inadequate data analysis
technique. Different effects
of non-circular beam have been discussed by different authors in \cite{Fosalba-2002,Mitra-2006}.
However, this particular effect of leakage from the dominant dipole has not been analyzed yet.

In this paper, analytical methods have been developed to calculate
the amount of power leakage from dipole and simulations with the actual scan pattern
of WMAP have been carried out showing the order of power leakage for
different WMAP beams. Our results show that the amount may
not be sufficient to explain the anomalies, but the power transfer
does have a measurable effect on the quadrupole. In anticipation, a similar
simulation has also been carried out with Planck scan pattern. As
the data for Planck beams are not available publicly we provide only upper limits
 to the beam non-circularity parameters beyond which the dipole leakage would cause detectable effect
in the power spectrum.

\section{Analytical description of the beam convolution}
This section describes the formalism employed for measuring the power leakage
from dipole to the quadrupole and higher multipoles during scanning
the sky with a non-circular beam. The measured sky temperature in
a CMB experiment is a convolution of the true sky temperature with
the beam function. If the measured temperature along $\gamma_{i}$ is expressed
by $\tilde{T}(\gamma_{i})$, where as the sky temperature along $\gamma$
is $T(\gamma)$ then 

\begin{equation}
\tilde{T}(\gamma_{i})=\int B(\gamma_{i},\gamma)T(\gamma)d\Omega_{\gamma}+T^{N}(\gamma_{i})\; .
\end{equation}

 Here $T^{N}$ is the noise in the scan procedure. Since we deal with low multipoles, the noise is ignored in our analysis. 
Here the beam function $B(\gamma_{i},\gamma)$ represents the sensitivity of the telescope around the pointing direction
$\gamma_{i}$.  This is a two point function and can be expanded in
terms of spherical harmonics as 

\begin{equation}
B(\gamma_{i},\gamma)=\sum_{l=0}^{\infty}\sum_{m=-l}^{l}b_{lm}(\gamma_{i})Y_{l}^{m}(\gamma)\; .
\end{equation}

Since we intend to measure the power leakage from dipole
to the quadrupole and it is convenient to consider only the sky dipole map
 and check the amount of power leakage from the dipole to higher multipole due to 
the non-circular beam. A dipole map $T(\gamma)$ can be written as a sum of
all the spherical harmonics with $l=1$, i.e. $T(\gamma)=\sum_{m=-1}^{1}a_{1m}Y_{1}^{m}(\gamma)$.
However, its always possible to choose a coordinate system such that
$a_{1,1}$ and $a_{1,-1}$ modes vanish and the sky temperature can
be expressed only as $T(\gamma)=T_{0}Y_{1}^{0}(\gamma)$, where $T_{0}=a_{1,0}$
is a constant. In such a case the measured sky temperature along the
$\gamma_{i}$ direction can be expressed
as 

\begin{eqnarray}
\tilde{T}(\gamma_{i}) & = & \int B(\gamma_{i},\gamma)T(\gamma)d\Omega_{\gamma}=\int\left[\sum_{l=0}^{\infty}\sum_{m=-l}^{l}b_{lm}(\gamma_{i})Y_{l}^{m}(\gamma)\right]T(\gamma)d\Omega_{\gamma}\nonumber \\
 & = & T_{0}\sum_{l=0}^{\infty}\sum_{m=-l}^{l}b_{lm}(\gamma_{i})\int Y_{l}^{m}(\gamma)Y_{1}^{0}(\gamma)d\Omega_{\gamma} =  T_{0}b_{10}(\gamma_{i})\label{eq:scanned temperature}\; .
\end{eqnarray}

The above expression can not be directly used for measuring the sky
temperature because it contains the beam harmonic coefficient
$b_{10}(\gamma_{i})$ , which is a function of $\gamma_{i}$. It is convenient
to orient the beam along some fixed direction of the sky, say along
the $\hat{z}$ direction, and consider the multipole $b_{lm}(\hat{z})$ to characterise the beam \cite{Souradeep-2001}. The spherical harmonic coefficients of the beam $b_{lm}(\gamma_{i})$ at any particular
direction $\gamma_{i}$ can obtained by using Wigner-D functions as  
\begin{eqnarray}
b_{10}(\gamma_{i}) & = & \sum_{m'=-l}^{l}b_{1m'}(z)D_{0m'}^{1}\nonumber \\
 & = & b_{1,-1}(z)D_{0,-1}^{1}(\varphi_{i},\theta_{i},\rho_{i})+b_{1,0}(z)D_{0,0}^{1}(\varphi_{i},\theta_{i},\rho_{i})+b_{1,1}(z)D_{0,1}^{1}(\varphi_{i},\theta_{i},\rho_{i})\nonumber \\
 & = & b_{1,-1}(z)d_{0,-1}^{1}(\theta_{i})e^{i\rho_{i}}+b_{1,0}(z)d_{0,0}^{1}(\theta_{i})+b_{1,1}(z)d_{0,1}^{1}(\theta_{i})e^{-i\rho_{i}}\label{eq:beam rotation}\; .
\end{eqnarray}

Substituting explicit expressions in terms of trigonometric functions for $d_{0,0}^{1}(\theta_{i})$
, $d_{0,-1}^{1}(\theta_{i})$ and $d_{0,1}^{1}(\theta_{i})$, in the
above and using eq(\ref{eq:scanned temperature}),
the expression for the scanned temperature can be written as 

\begin{equation}
\tilde{T}(\gamma_{i})=T_{0}b_{10}(z)\cos(\theta_{i})+\sqrt{2}T_{0}\sin(\theta_{i})\left[b_{r}(z)\cos(\rho_{i})+b_{i}(z)\sin(\rho_{i})\right]\label{eq:temp}\; .
\end{equation}

Here $\rho_{i}$ is the orientation of the semi-major axis of the
beam at the $i^{th}$ scan point. The function $b_{i}(z)$ and $b_{r}(z)$
can be defined as follows. In eq(\ref{eq:beam rotation}) $b_{1,1}$
and $b_{1,-1}$ are complex quantities. Since the beam is real,
the beam spherical harmonic coefficients should satisfy the 
relation $b_{1,1}^{*}=-b_{1,-1}$. Here, $b_{r}$ and $b_{i}$ has been defined as $b_{1,1}=b_{r}+ib_{i}$,
i.e. the real and the imaginary part of $b_{1,1}$.

From eq(\ref{eq:temp}) we can calculate the power that gets
leaked to the quadrupole or the higher multipoles. It can be seen
from eq(\ref{eq:temp}) that the term
with $b_{10}(z)$ will not contribute to any power leakage from dipole.
Therefore the dipole to quadrupole power transfer is caused only by
the terms multiplied with $b_{r}(z)$ or $b_{i}(z)$. Hence, if the experimental beam is designed in such a
way that the $b_{r}(z)$ or $b_{i}(z)$ components of the beam are
completely negligible than it is possible to completely eliminate the dipole to higher multipole power
transfers.

\section{Dipole leakage from WMAP scan pattern}

 The WMAP satellite follows a unique scan pattern, in which pixels near the two
poles are scanned for large number of times,
whereas those which are near the equator are scanned for least number of times. 
The satellite scans the sky temperature in five different frequency
bands, named as $K$, $Ka$, $Q$, $V$ and $W$ in a differential measurements 
of a pair of horns. Amongst them $Q$
and $V$ band have two detectors each and $W$ band has four detectors.
Each of these detectors has a pair of horns, both of which are are about $~70.5^{o}$
off the symmetry axis. It has a fast spin about the symmetry
axis with the spin period of around $2.2$ minutes. Along with this
fast spin, the spacecraft has a slow precession, $22.5^{o}$
about the Sun-WMAP line. This precession period is about $1$ hour.
The earth-sun vector rotates $360^{o}$ in a year. 

The simulation has been carried with a dipole map, similar to the known CMB
dipole. We assumed that the shape
of the two beams for a pair of horns of a detector are almost same, i.e.
 the parameters $b_{r}$, $b_{i}$ and $b_{10}$ are
same for both the beams of a detector. 

All the simulations have been carried out on Healpix map with $N_{side}=256$ resolution.
The simulation gives us three independent maps from the three independent
components of eq(\ref{eq:temp}). The figures from the three maps
are shown in the Fig.\ref{fig:WMAP-simulation}. All these
maps are shown in ecliptic coordinate
system. These maps can be multiplied with $b_{10}$, $b_{r}$ and
$b_{i}$ and then summed up to get the final scanned map from a detector
and hence the amount of power transfer can be calculated.  The values of the beam
spherical harmonic coefficients i.e. $b_{r}$, $b_{i}$ and $b_{10}$
along with the amount of power leakage are listed
in the table \ref{tab:The-coefficients}. 

The analysis shows that the temperature leaked into the quadrupole from the dipole is
less than $2\mu K$ compared to the CMB quadrupole measured by WMAP of $\sim6\mu K$.
Although the dipole power leakage is small, the amount of power leakage is
sufficient to suggest a  deeper analysis of the WMAP data for this effect.

\begin{table}
\caption{\label{tab:The-coefficients}The dipole coefficients,  $b_{r}$ and $b_{i}$ (real and 
imagionary part of $b_{1,1}$) of the beam spherical harmonics for different WMAP beams 
estimated from the publicly available beam maps \cite{beam}. The quadrupole and octapole temperature are calculated from the simulation considering the dipole temperature as 3.358 mK. }
\begin{center}
\lineup
\begin{tabular}{lllllll}
\br 
 & $b_{r}$ & $b_{i}$ & $T_{d}/T_{q}$ & $T_{q}(\mu K)$ & $T_{d}/T_{oc}$&$T_{oc}(\mu K)$ \tabularnewline
\mr
$K_{1}$ & $\-1.45 \times 10^{-4}$ & $2.37 \times 10^{-5}$ & $\01837.9$ & $1.82$ & $\011250.6$ & 0.30\\ 
$Ka_{1}$ & $\-9.42 \times 10^{-5}$ & $\-6.03 \times 10^{-5}$ & $\02568.3$ & $1.31$ & $\016493.7$ & 0.20\\
$Q_{1}$ & $8.66 \times 10^{-5}$ & $\-1.62 \times 10^{-4}$ & $\02748.1$ & $1.22$ & $\014226.4$ & 0.24\\
$V_{2}$ & $7.40 \times 10^{-5}$ & $8.11 \times 10^{-5}$ & $\03256.4$ & $1.03$ & $\019674.1$ & 0.17\\
$W_{4}$ & $6.55 \times 10^{-6}$ & $2.23 \times 10^{-5}$ & $33048.57$ & $0.10$ & $133621.8$ & 0.025\\
\br
\end{tabular}
\end{center}
\end{table}

\section{Dipole leakage from Planck scan pattern}

The Planck satellite has only one beam for each detector and thus
instead of the differential measurement it measures the actual temperature
of the sky. The Planck satellite beam is approximately $85^{o}$ off
symmetric axis and the precession angle for the satellite is around
$7.5^{o}$. The precession rate is taken as one revolution per six
month and the spin rate as $180^{o}/min$. 

\begin{figure}[h]
\begin{minipage}{18pc}
\includegraphics[width=0.32\columnwidth]{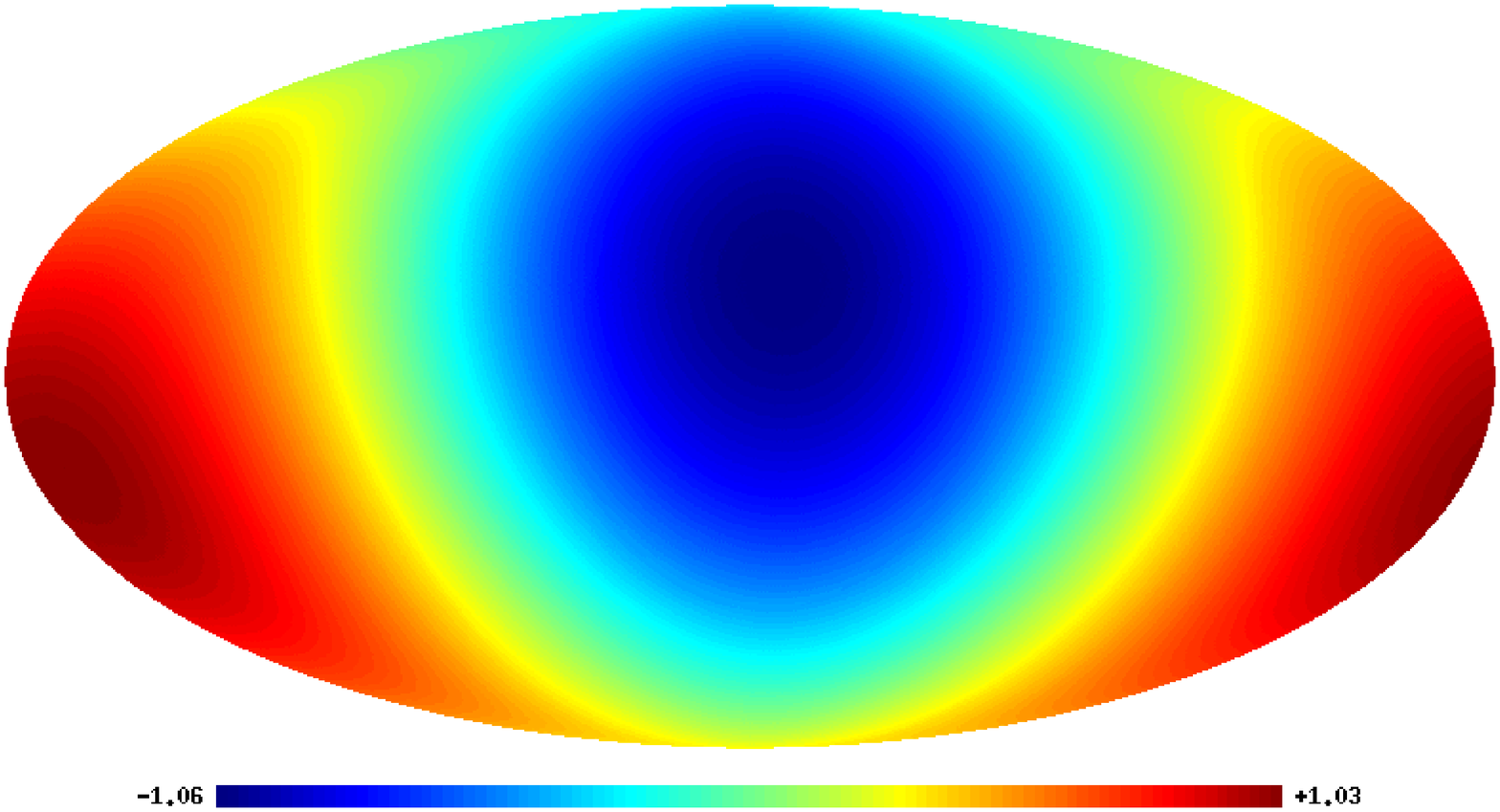}\includegraphics[width=0.32\columnwidth]{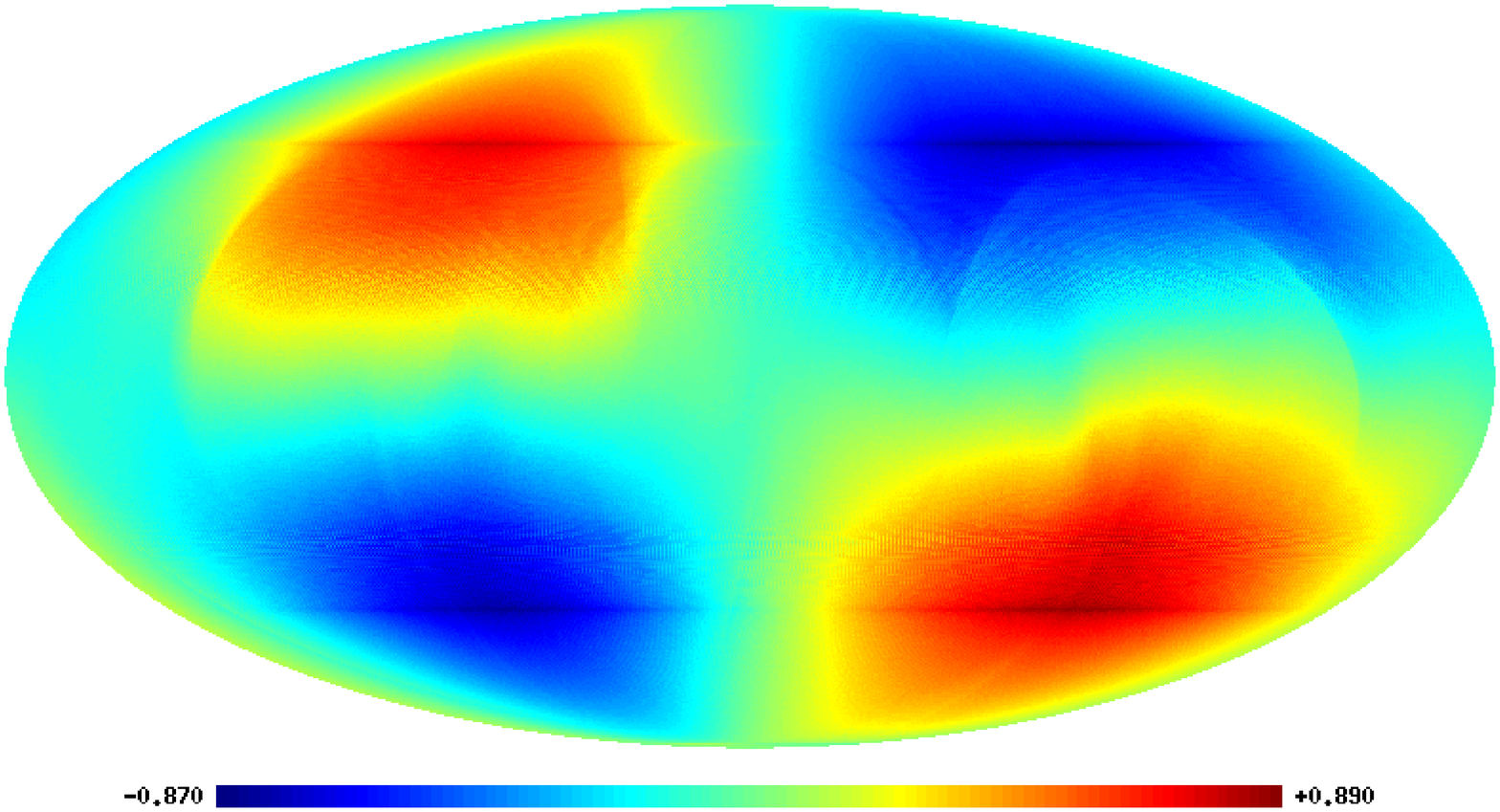}\includegraphics[width=0.32\columnwidth]{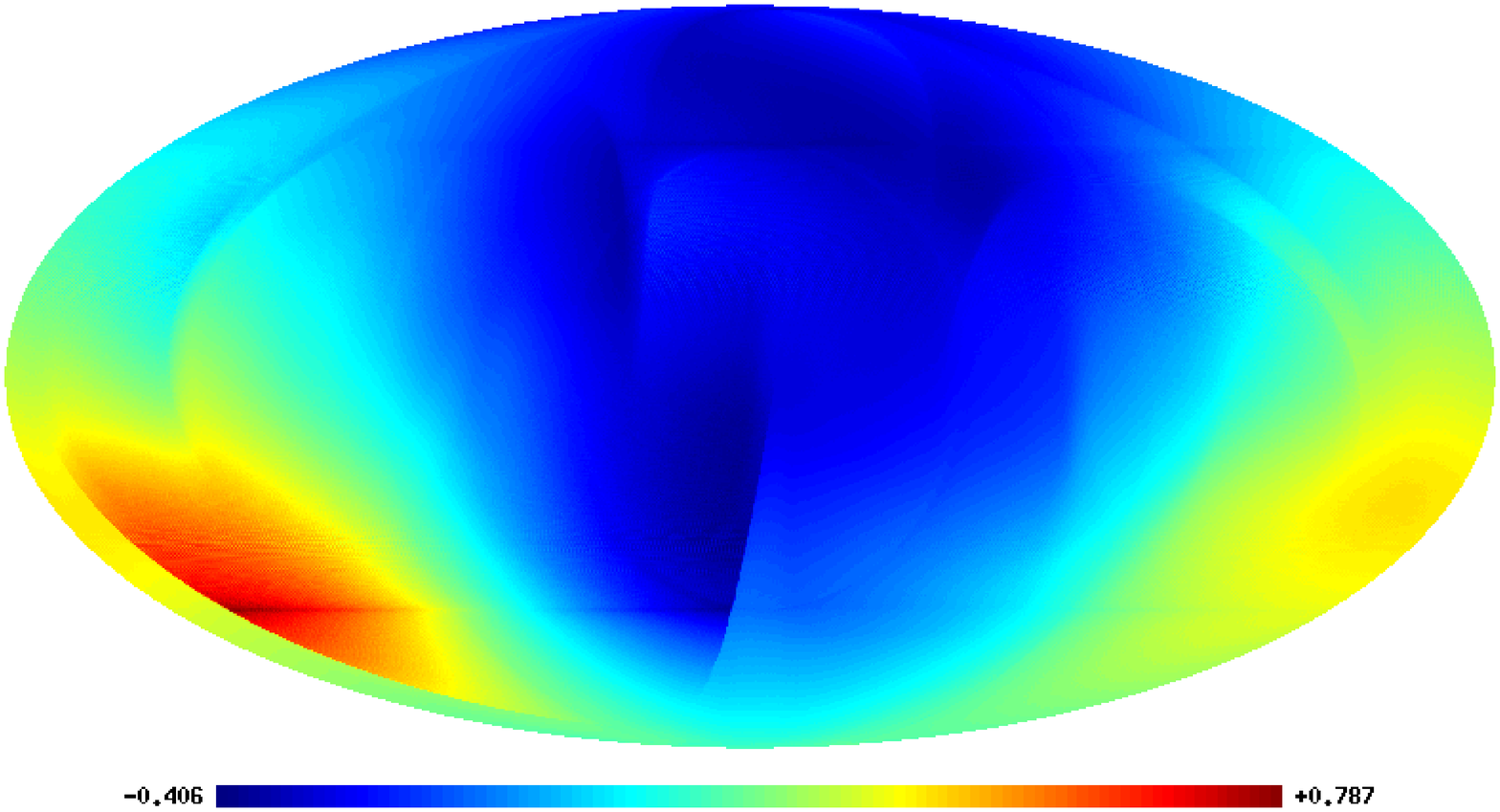}

\caption{\label{fig:WMAP-simulation}Simulated maps of $\cos\theta$, $\sin\theta\cos\rho$ 
and $\sin\theta\sin\rho$ component from WMAP scan pattern in the Ecliptic coordinate system. }
\end{minipage}\hspace{2pc}%
\begin{minipage}{18pc}
\includegraphics[width=0.32\columnwidth]{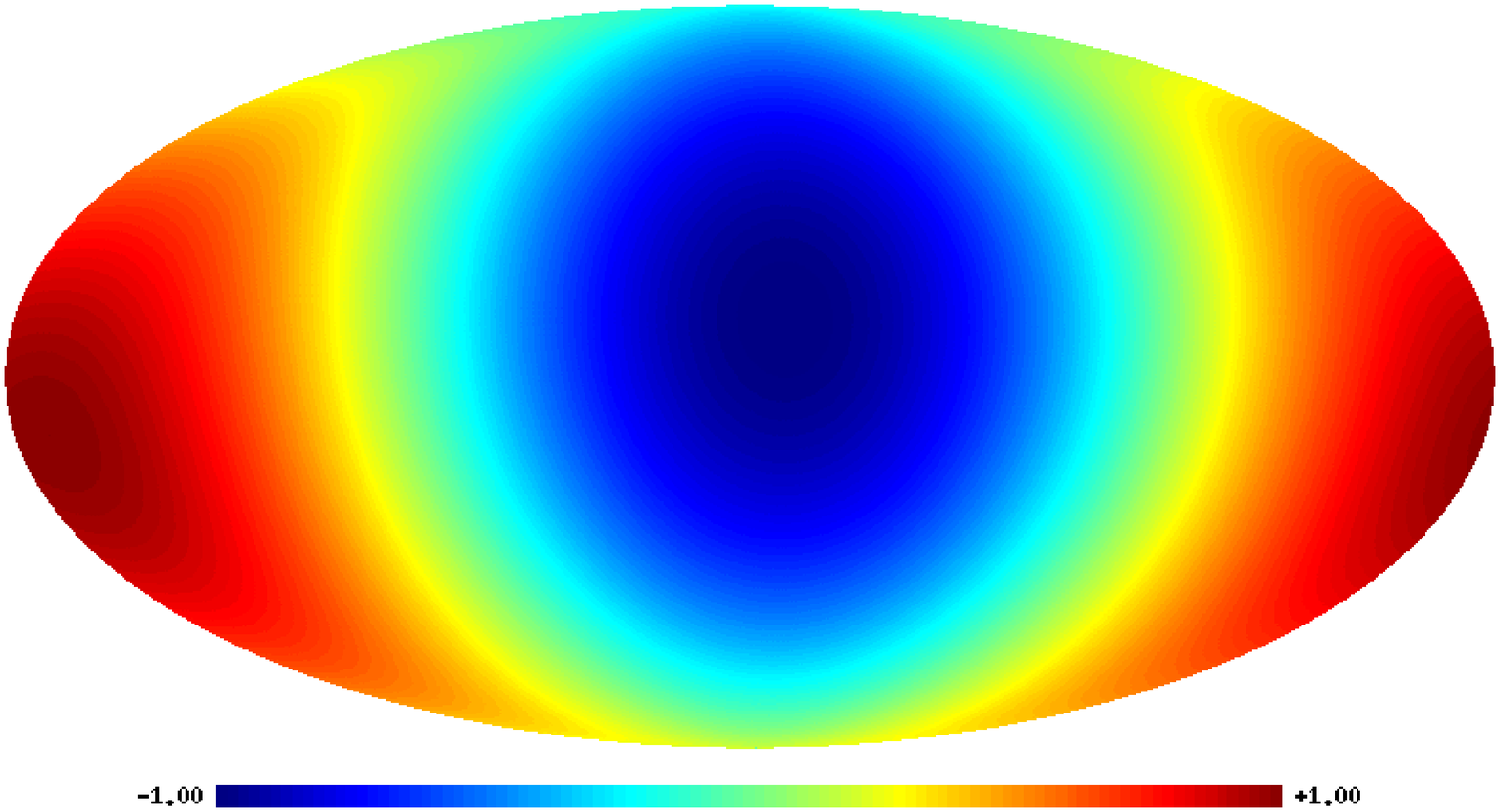}\includegraphics[width=0.32\columnwidth]{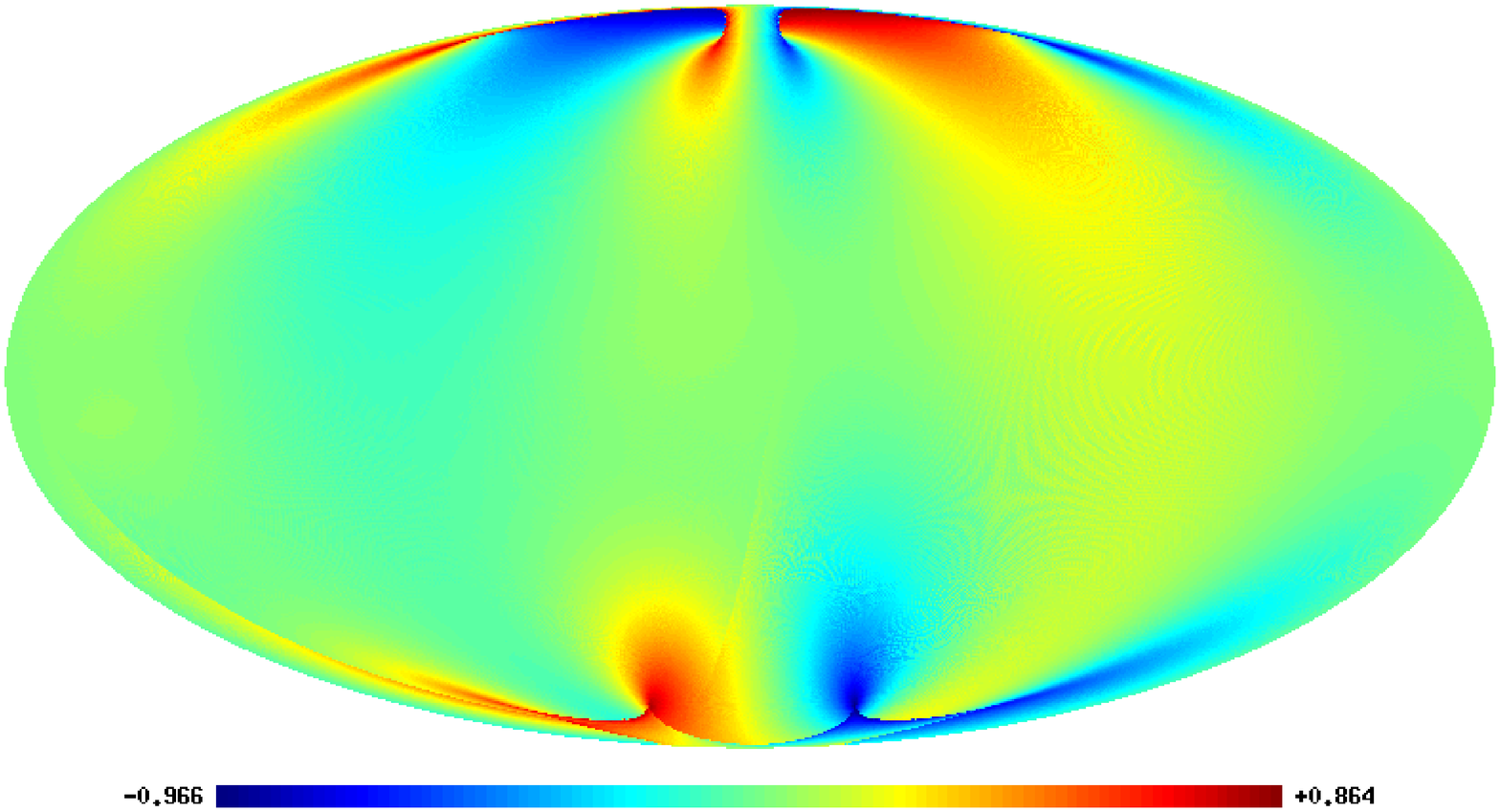}\includegraphics[width=0.32\columnwidth]{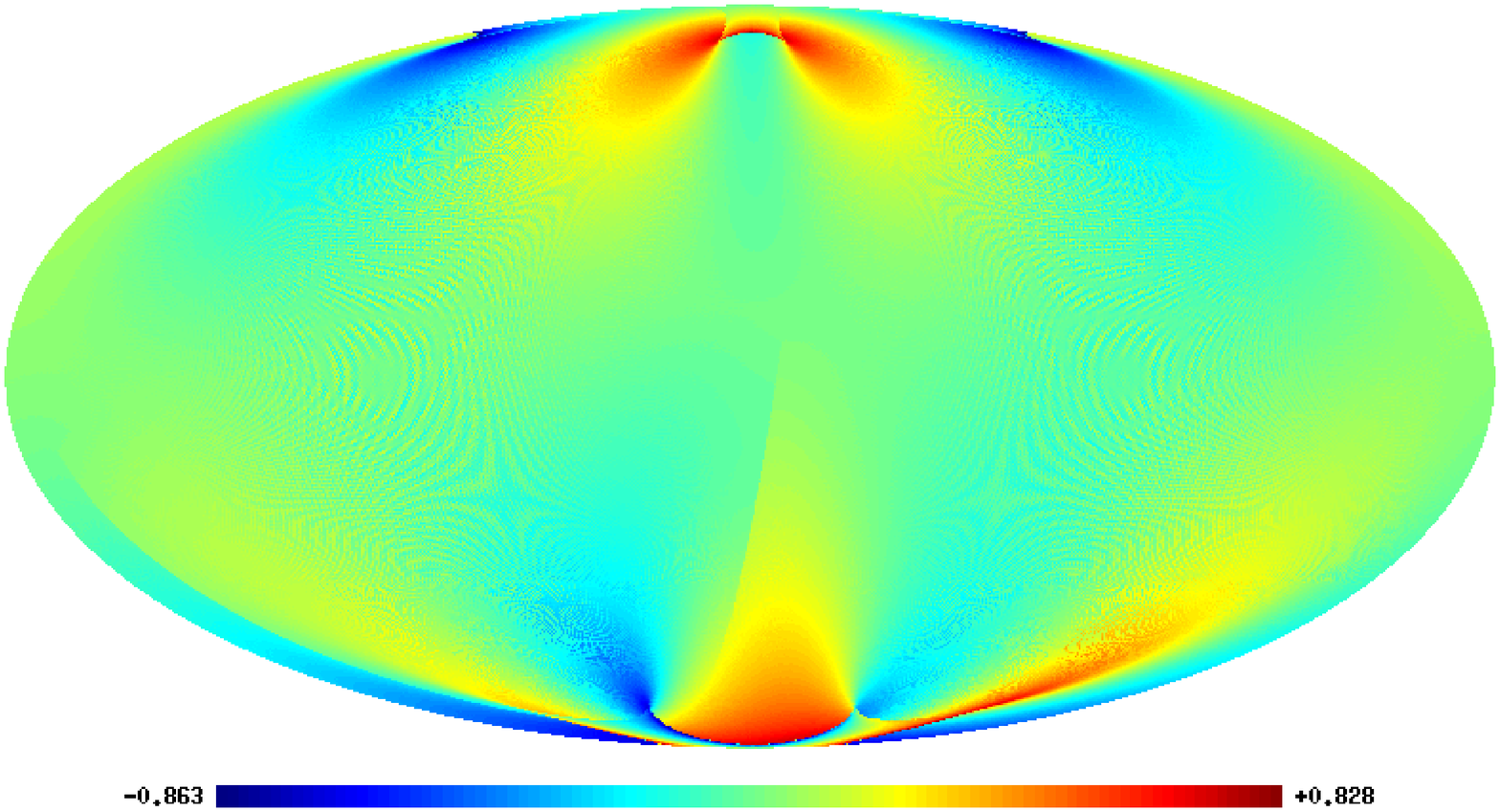}

\caption{\label{fig:Planck-simulation}Simulated maps of $\cos\theta$, $\sin\theta\cos\rho$ 
and $\sin\theta\sin\rho$ component from Planck scan pattern in Ecliptic coordinate system. }
\end{minipage} 
\end{figure}

An analysis similar to that of WMAP has been carried out for Planck
satellite also. Three independent maps are calculated from the
Planck scan strategy are shown in Fig.\ref{fig:Planck-simulation}.
As the data of the Planck beam map is not publicly available therefore the
amount of power leakage expected from the dipole can not be computed for the
Planck scan pattern. But, the analysis shows that if the $b_{i}$ and $b_{r}$
are of the order of $5\times10^{-4}$ or less then it will cause a
temperature leakage less than $<1\mu K$. Therefore, that amount of
leakage may be safely ignored for beams satisfying the above limits. 

\section{Conclusion}

An analytical formalism has been developed to use in simulations of scan strategy 
to estimate the leakage of power in the dipole anisotropy to the quadrupole
and higher multipoles. It has also been shown that the power leakage
only depends on the two spherical harmonic coefficients of the satellite
beam ($b_{i}$ and $b_{r}$) and therefore if the beam has been designed
in such a way that these two parameters of the beam are small enough then power 
leakage will be negligible. For WMAP, the amount of the power leakage
is found to be small but not insignificant compared to the low value of quadrupole measured. 
The simulations also show that the amount of power
leakage depends on the scan pattern. For identical beam shapes,
the amount of power leakage is more in WMAP scan strategy compared to Planck scan strategy.

\end{document}